\documentclass{svproc}
\usepackage[utf8]{inputenc}

\usepackage{url}

\usepackage[sectionbib]{natbib}
\bibpunct{(}{)}{;}{a}{}{,}%
\usepackage{graphicx}
\usepackage{amsmath,amssymb}
\usepackage{multirow}
\usepackage{longtable}
\usepackage{changes}

\usepackage[b5paper]{geometry}
\begin{document}
	\mainmatter              
	\title{Caucasian Mountain Observatory of Sternberg Astronomical Institute: first six years of operation }
	\titlerunning{First six years of CMO SAI MSU}  
	%
	
	\author{Nicolai Shatsky \and Alexander Belinski \and Alexander Dodin \and Serguey Zheltoukhov \and Victor Kornilov \and Konstantin Postnov\and Serguey Potanin \and Boris Safonov \and Andrei Tatarnikov \and Anatol Cherepashchuk }
	\authorrunning{Nicolai Shatsky et al.} 
	%
	\tocauthor{N. Shatsky, A. Belinski, A. Dodin et al.}
	\institute{Lomonosov Moscow State University, Sternberg Astronomical Institute, Universitetskij pr. 13, Moscow 119234, RUSSIA  \\
		\email{kolja@sai.msu.ru},
	\\ WWW home page:
		\texttt{http://www.sai.msu.ru}
		}
	
	\maketitle

\begin{abstract}
The new SAI MSU observatory 2.5-meter telescope and capabilities of its current instrumentation are described. The facility operates actively since 2014 in parallel to the engineering works. It has delivered a number of prominent results in the field of optical and near-infrared photometry and spectroscopy as well as newly developed observational techniques like differential speckle polarimetry.
\end{abstract}

\section{CMO SAI MSU: fact sheet}

The Caucasian Mountain Observatory of Sternberg Astronomical Institute (CMO SAI MSU) was constructed in 2011-2013. It is located at an altitude of 2100 m asl at the Shadjatmaz plateau 20 km south to the city of Kislovodsk at Northern Caucasus (coordinates: 43$^o$44'10"N, 42$^o$40'03"E). The astroclimatic conditions of the site are reported in \cite{2014PASP..126..482K}. The observatory instrumentation includes a 2.5-m Ritchey-Chretien alt-azimuth reflector, a 60-cm robotic RC600 reflector, and an astro-climate monitor. The first light from the 2.5-m telescope was received on November 11, 2014. Since then it has been operating interchangeably in the test (engineer) and scientific observation mode. The RC600 telescope is continuously operating in the remote mode since May 2019.

\section{2.5-m telescope of CMO SAI MSU}

The main instrument of CMO SAI is a 2.5-meter Ritchey-Chretien reflector built for scientific and educational works to be performed by the University staff, students and postdocs. It was designed and built by a consortium headed by Safran REOSC (France; general design and optics manufacturing), NIAOT (Nanjing Institute of Astronomical Optics and Technology, China; the mount), Gambato SAS (Italy, the dome) and SAI--Servotechnica LLC partnership (the new control system), with the overall provision of the project run by the Maveg Industrieaurustungen, GmbH (Germany).

The SAI2.5 is a universal multiport (Cassegrain, 2 Nasmyths plus 2 ``student'' Nasmyth ports) alt-az F/8 telescope having 40,000 sq.cm clear aperture and equipped with three main port rotators and two off-axis CCD autoguiders. It possesses the classic (rigid) zero-expansion substrate optics capable of delivering the images with 80\% of encircled energy within $r=0\farcs3$ \citep{2017ARep...61..715P}; the actual atmosphere-impacted image quality happens to be as good as $FWHM=0\farcs5$ at the Cassegrain focus. The uncorrected field of view (10\arcmin) may be expanded up to 40\arcmin\ using a three-lens silica Wynne corrector.

The telescope is operating at slew rates $3^o/s$ and blind points with a precision of around 2\arcsec\ RMS. Tracking precision of 0.1\arcsec\ RMS is maintained up to the wind speeds 10~m/s. The control hardware is based on the Etel direct drives and Kollmorgen BLDC motors driven by the DeltaTau PMAC2 motion controller (main axes); the auxiliary units (mirrors, guider arms etc) control system consists of distributed Nanotec drives.

The 2.5-m telescope is operating by a software developed at SAI. Its component units (\emph{main axes} and \emph{auxiliary units} of the mount, the \emph{dome}, its \emph{climate-control} and \emph{weather station}; the telescope \emph{interface for scientific equipment} and the databases for \emph{observation results metadata} and \emph{programs and targets}) are operated via the EPICS bus by two chief program: the \emph{Observations control software} and the \emph{Object planner}. This structure enables the flexible scheduling of instruments, objects and programs depending on the current weather and seeing conditions continuously provided by the Astroclimatic monitor (ASM, \citealt{2010MNRAS.408.1233K}).

The telescope is currently equipped with the following full-time instrumentation:
\begin{description}
\item[ASTRONIRCAM] -- imaging IR low-res spectrometer (Nasmyth-1, since 2015)
\item[NBI] -- the Niels Bohr institute wide field CCD imager (Cass-1, 2014)
\item[TDS] -- Transient double-beam optical low-resolution spectrograph (Cass-2, 2019)
\item[Speckle Polarimeter] -- high angular resolution multi-mode instrument (Nasmyth-2, 2015)
\end{description}

\subsection{NBI camera}
The Wide Field Imager (NBI 4Kx4K camera) is based on two EEV CCD4482 detectors operating at $-120^o$C with an LN2 cryostat and works with the Filter \& Shutter Unit hosting a wide set of filters for the $10\arcmin\times10\arcmin$ field of view: $UBVRI$ Bessel set, SDSS set and narrow-band H$\alpha$, H$\beta$, [OIII] \& [SII] filters with continua. This is a first light instrument of SAI2.5 \citep{2015IBVS.6126....1A} built by NBI, Copenhagen University.

As a simple and efficient CCD imager, it reaches the sky background level objects ($g',r'\!=\!21.8^m$) with a 300~s integration having signal to noise SNR=20.

\subsection{ASTRONIRCAM}
The Astronomical Near-Infrared Camera-spectrograph is a cryogenic (T=77K) focal reducer based on the Hawaii-2RG 2Kx2K detector built by the Mauna Kea Infrared LLC. It operates in the photometry and the low resolving power (R=1200) spectroscopy modes (see \citealt{2017AstBu..72..349N} for a detailed description).

The filter set includes the MKO system $J$, $H$, $K$ and $Ks$ wide-band filters and narrow-band filters for the $CO$, $H2$, $[FeII]$ and Br$\gamma$ lines. The objects of $J>8^m$ magnitude (saturation limit set by the minimal exposure time 1.8~s) may be measured in the $4\farcm6\times4\farcm6$ field in the photometric mode. An accurate account of various error sources is performed via the \emph{dithered} mode of telescope acquisition and may reach the $J=20^m$, $K=18^m$ sources with a 10\% precision in 1000~s integration.

The long-slit ($Y$, $J$, $H$, $K$) and cross-dispersed ($YJ$ and $HK$; 10\arcsec\ slit) spectroscopy modes are available for objects brighter than $15^m$ \citep{Zheltoukhov_anc2020}. The efficiency of these modes is low (6--14\% and 1--2\%) so it is planned replace the non-efficient grisms installed up to now. 

\subsection{TDS}
The Transient Double-beam Spectrograph is a new SAI2.5 instrument built for the high-throughput low-resolution optical spectroscopy of non-stationary and extragalactic sources. Installed at the side Cassegrain port, it is aimed to serve as an always-ready tool for observations of non-stationary objects and transients.  

The TDS works with 1\arcsec\ and 10\arcsec\ wide, 3\arcmin\ long slits and consists of two arms with the VPH-gratings delivering 560--740~nm (R=2500) and 350--580~nm (R=1250) spectral domains to a couple of Andor Newton DU940P cameras; the light is split between the channels with a dichroic plate. The doubled resolution is also available in the 450--560~nm range. The high transparency of the optics and dispersers  results in an overall (slitless) efficiency of 30--45\% in the blue--red beams, respectively, which enables spectra of $g'=20^m$ to be recorded in 30 min with SNR=5 \citep{Dodin2020a}. The TDS control and data reduction software is developed in Python.

\subsection{Speckle polarimeter}
The polarimetry and high angular resolution tasks are tackled with the Speckle polarimeter (SPP, \citealt{Safonov2019a}) built at SAI in 2015. It is a double-beam Wollaston prism polarimeter with a rotating half-wave plate equipped with an EMCCD camera Andor iXon~897. By imaging a narrow $5\arcsec\times10\arcsec$ sky patch simultaneously in two linear polarizations with the sub-diffraction resolution and high frame rate, it is capable of working in the following modes: \emph{total flux polarimetry},  \emph{speckle interferometry}, \emph{differential speckle polarimetry} and \emph{fast photometry} or \emph{lucky imaging}. 

The instrument operates through standard Bessel $BVRI$ or middle-band $\lambda\lambda$ 550, 625 \& 880~nm filters and has an Atmospheric Dispersion Corrector.

\section{60-cm robotic RC600 telescope}
This D=600 F=4200~mm Ritchey-Chretien telescope (ASA Astrosysteme GmbH, Austria) was intended as a photometric ground-base support of space projects as well as SAI2.5 programs, a remotely controlled or robotic photometric facility for monitoring of various non-stationary objects and exoplanet transits. It started operating in May 2019 delivering a lot of observations of various SNs, CVs, Cepheids, AGNs, microquasars and exoplanets during the very first year.

The modern direct-drive DDM160 equatorial mount is capable of retaining the target to within an arcsecond-size box. The Andor iKon-L BV CCD camera has a wide $22\arcmin\times22\arcmin$  FoV and can take images in the Johnson/Cousins  $UBVRcIc$ and Sloan $g'$, $r'$ and $i'$ filters; the practically measurable magnitude of the photometry for 60s exposures (SNR=5) is $Rc\sim19.3^m$ \citep{Berdnikov20-4}.

\section{Astroclimatic monitor}
The ASM facility is an oldest one at CMO and  operates continuously since the end of 2007. It has collected a long series of atmospheric turbulence profile measurements together with the sky background and extinction and the site weather conditions data. In 2020, the ASM started to feed the flexible target scheduler software for SAI2.5 operators. In 2018--2020, the weather statistics implies a median seeing of 1\farcs02 and 1300--1400~hr of clear sky per year.

\section{Highlight results of CMO SAI MSU}
In the recent years, the CMO  instruments were used to obtain results in 
several topical fields of research, including:
\begin{itemize}
    \item  Variability and size-luminosity relation of AGNs \citep{Oknyansky_anc2017, ilic2020, Cho_anc2020}
    \item Study of variable stars associated with maser sources \citep{Sobolev_anc2020}
    \item  Star-forming regions at the periphery of the supershell surrounding the Cyg OB1 association \citep{Sitnik_anc2020, Sitnik_anc2019, Tatarnikova_anc2016}
    \item Optical and NIR-photometry of X-ray sources with black holes, microquasars, pre-main sequence and post-AGB objects \citep{Cherepashchuk_anc2020, Cherepashchuk_anc2019}
    \item Spectroscopy of an ultramassive white dwarf \citep{Pshirkov2020}
    \item Colour and polarimetric variability of UXORs; young stellar object jets photometry \citep{Dodin_anc2020, Dodin_anc2019, Berdnikov_anc2017}
    \item Analysis of temporal evolution of the circumstellar envelopes of evolved stars at high angular resolution \citep{Safonov2019b, Fedoteva2020}
    \item Spectral characterization of solar system bodies \citep{Busarev_anc2018}
    \item Near‐infrared photometry of superthin edge‐on galaxies \citep{Bizyaev_anc2020}
    \item Spectral  characterization and identification of \textit{SRG/eRosita} sources \citep{Dodin2020a} 
    \item High-accuracy monitoring of exoplanet transits \citep{2020arXiv200911899B}.
\end{itemize}

In 2021, the 2.5-m reflector of CMO SAI is expected to take commissioning and start operating as part of the Russian large telescopes program.

\paragraph{Acknowledgements.} The construction and development of CMO SAI MSU have been supported by M.V. Lomonosov Moscow State University Program of Development, RSF grants  17-12-01241 and 16-12-10519, RFBR grants 16-32-60065, and the Leading Scientific School of MSU ``Physics of stars, galaxies and relativistic objects''. This work is supported by the Ministry of science and higher education under the contracts 075-15-2020-780 (exoplanet transits) and 075-15-2020-778 (observations of objects with extreme energy release) in the framework of the Large scientific projects program within the national project ``Science''.
%
%

\end{document}